\documentclass[a4paper,mtpluscal]{aipproc}

\def\del{\partial}
\layoutstyle{6x9}

\begin{document}

\title{The Nucleon and Roper Resonance in a Chiral Quark Diquark Model}




\author{Keitaro Nagata and Atsushi Hosaka}{
  address={Research Center for Nuclear Physics, Osaka University,
567-0047, Mihogaoka, Osaka, Japan},email={nagata@rcnp.osaka-u.ac.jp}
}

\begin{abstract}
A description of the nucleon and Roper resonance in  a
quark-diquark approach is presented. We show that two states with
the quantum number of the nucleon can appear in the ground state of
the spatial configuration, when there are two types of diquarks:
scalar-isoscalar and axial-vector-isovector diquarks. 
The mass difference between the two 
states is generated by the mass difference between the two diquarks, 
which is due to the spin-spin interaction
between the two quarks in the diquarks. The two states are then
identified with the nucleon and Roper resonance.
\end{abstract}

\maketitle


\section{Introduction}
Chiral symmetry with its spontaneous breaking is a powerful tool to 
investigate the low energy dynamics of QCD. 
The internal structure of hadrons is also an important ingredient, as
we investigate phenomena at finite momentum transfer.
The Nambu-Jona-Lasinio
(NJL) model is one of the effective theories inspired by chiral
symmetry\cite{Nambu:1961fr}, 
and has been used to study the dynamics of mesons from the vacuum to
finite density/temperature system. 
A chiral quark-diquark model is an effective approach, which is an
extension of the NJL model for the study of baryons by the inclusion of the diquark 
degrees of freedom. This model
incorporates not only the mesons but also baryons as composite
particles with respecting chiral symmetry, where mesons and baryons
are quark-antiquark and quark-diquark bound states, respectively. 
It was used to study not only one particle properties of the
nucleon\cite{Ebert:1997hr,Abu-Raddad:2002pw,Nagata:2004ky}, but also
meson-baryon\cite{Ebert:1997hr,Abu-Raddad:2002pw} and baryon-baryon
interactions\cite{Nagata:2003gg}.
The importance of the diquark correlation has been also suggested from
the recent study of exotics\cite{Jaffe:2003sg,Jaffe:2004ph}. 
Current authors suggested the possibility that the Roper
resonance is described together with the nucleon by 
the mixing of two types of quark-diquark channels\cite{Nagata:2005qb}, 
where the correlation between quarks plays an important role. 

It was shown that two local operators can be independently used for
the study of the nucleon\cite{Ioffe:1981kw}. Inspired by this, here
we prepare a model that can generate two types of nucleon states from two
independent diquarks: the scalar and axial-vector diquarks.
In naive quark models(NQM) with uncorrelated quarks 
in the $(0s)^3$ configuration, one of the two states is forbidden due to
spin-flavor $SU(6)$ symmetry. Therefore, in NQMs the second state with
the quantum number of the nucleon $I(J)^P=1/2(1/2)^+$, which is known
as the Roper resonance, must be a radially excited state with $N=2$ and
the excitation energy  $2\hbar\omega$. If, however, the diquark
correlation becomes significant, two states appear as active degrees
of freedom, generating the two nucleon states approximately  as bound
states of a quark and a diquark. This is the case we consider in the
present paper, where we identify the higher state with the Roper resonance.
In this case the mass difference of the nucleon and Roper is approximately
dictated by the mass difference of the two diquarks, the origin of which is the
spin-spin interaction between quarks, and is about the same order as the
mass splitting of the nucleon and delta.

\section{Framework}
We employ the chiral quark-diquark
model\cite{Abu-Raddad:2002pw,Nagata:2005qb}, which is given by 
\begin{eqnarray}
{\cal l} = \bar{\chi}_c(i\rlap/\del - m_q) \chi_c \;+\;
D^\dag_c (\del^2 + M_S^2)D_c
 +
{\vec{D}^{\dag\;\mu}}_c 
\left[  (\del^2 + M_A^2)g_{\mu \nu} - \del_\mu \del_\nu\right]
\vec{D}^{\nu}_c +L_{int},
\label{lsemibos}
\end{eqnarray}
where $\chi_c$, $D_c$ and $\vec{D}_{\mu c}$ are the constituent quark, scalar
diquark and axial-vector diquark fields, and $m_q$, $M_S$ and $M_A$
are the masses of them. The indices $c$ represent the color. The term $L_{int}$ is the quark-diquark interaction, which is written as
\begin{eqnarray}
L_{int}&=&G_S\bar{\chi}_cD^\dagger_c
D_{c^\prime}\chi_{c^\prime}+v(\bar{\chi}_cD^\dagger_c\gamma^\mu\gamma^5
\vec{\tau}\cdot\vec{D}_{\mu c^\prime} \chi_{c^\prime}+\bar{\chi}_c\gamma^\mu\gamma^5
\vec{\tau}\cdot\vec{D}^\dagger_{\mu c}
D_{c^\prime}\chi_{c^\prime})\\
&+&G_A\bar{\chi}_c\gamma^\mu\gamma^5
\vec{\tau}\cdot\vec{D}^\dagger_{\mu c}
\vec{\tau}\cdot\vec{D}_{\nu c^\prime}\gamma^\nu\gamma^5 \chi_{c^\prime},
\label{eq:twoc}
\end{eqnarray}
where $G_S$ and $G_A$ are the coupling constants for the quark and scalar
diquark, and for the quark and axial-vector diquark. The coupling
constant $v$ causes the mixing between the scalar and axial-vector
channels. It is important that there are
two quark-diquark channels and they are independent. Hence the 
nucleon and Roper states appear as physical states. 
An effective Lagrangian for them
 is then derived by the path-integral hadronization method
\cite{Abu-Raddad:2002pw}.

\section{Masses of two states}
\begin{figure}
\includegraphics[width=10cm]{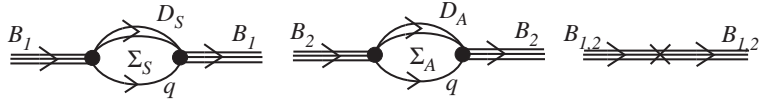}
\caption{A diagrammatic representation of the quark-diquark
   self-energy for the scalar diquark(left), axial-vector
diquark(middle) and mixing between them(right). 
The single, double and triple lines represent the quark,  
   diquark and nucleon respectively. The blobs represent the three point  
   quark-diquark-baryon interactions.}\label{fig:self}
\end{figure}

\begin{figure}
  \includegraphics[width=5cm]{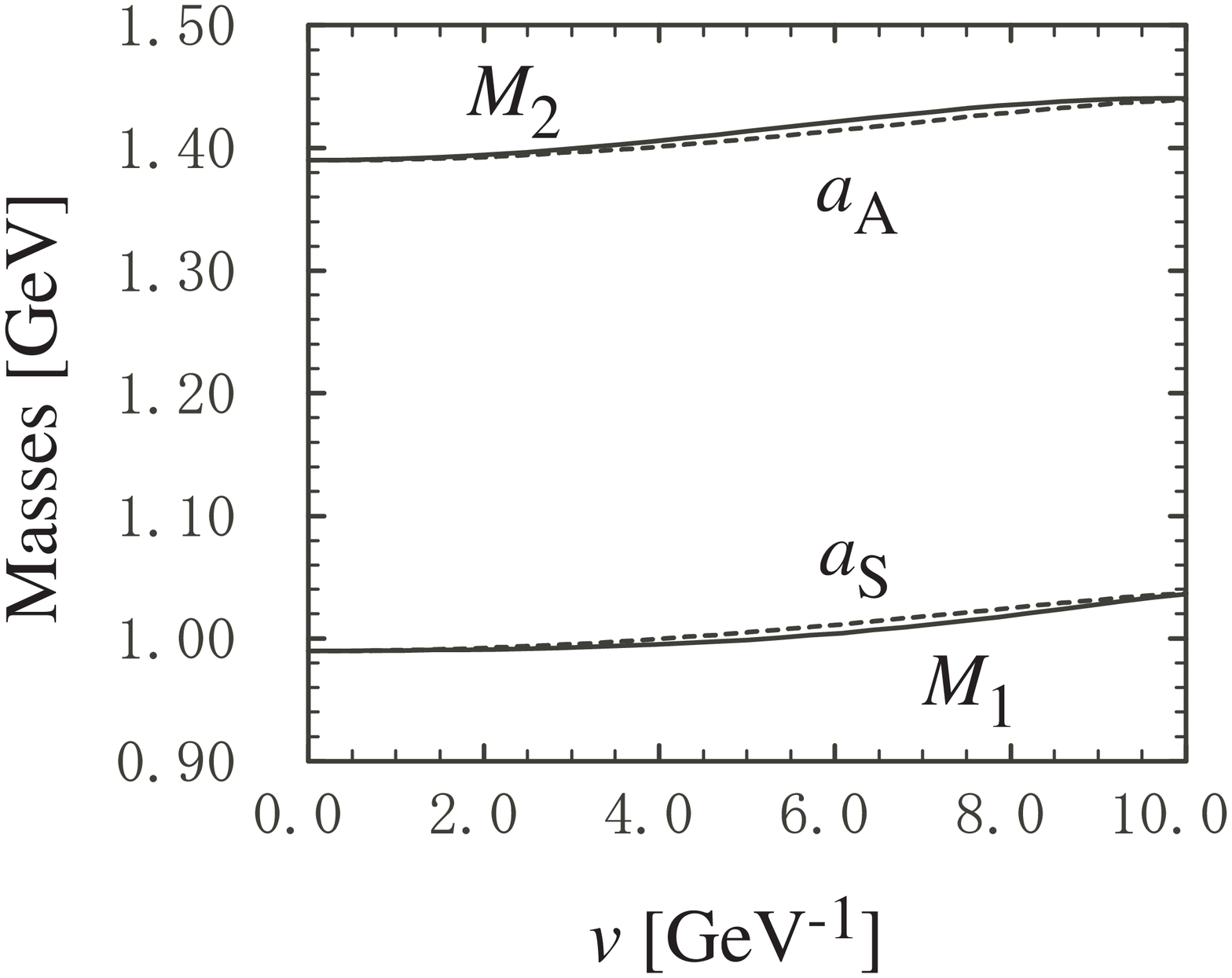}
  \includegraphics[width=5cm]{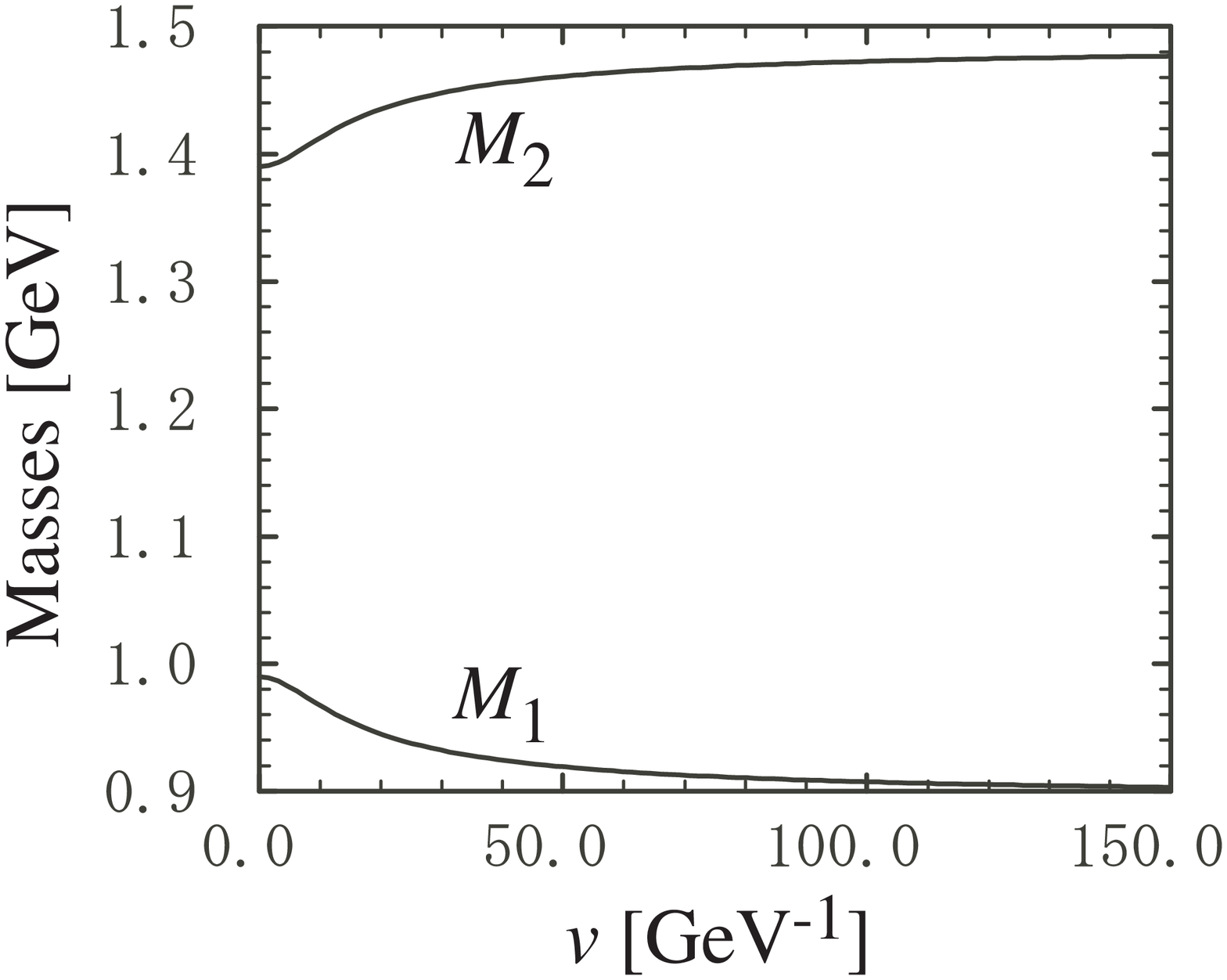}
  \caption{$M_1$ and $M_2$ as functions of $v$ for fixed $G_1$ and
  $G_2$(left) and for fixed $a_S$ and $a_A$(right).}\label{fig:massvsmixing}
\end{figure}

In this work we concentrate on the masses of the two nucleon states,
which are obtained from the quark-diquark self-energies shown in
figure \ref{fig:self}. After the calculation of the self-energies and
diagonalization of the $2\times 2$ mass matrix for the scalar and
axial-vector channels, the masses of physical states 
are obtained as follows\cite{Nagata:2005qb},
\begin{eqnarray}
M_{1,2}=\frac12\left[a_S+a_A\pm
\sqrt{(a_S-a_A)^2+4Z_S Z_A
\left(\frac{v}{|\hat{G}|}\right)^2}\right]\, , \\
\end{eqnarray}
where $a_S$ and $Z_S$ are the mass and wave-function renormalization
constant of the self-energy for the scalar channel, $a_A$ and $Z_A$ are those
for the axial-vector channel, obtained by
\begin{eqnarray}
\Sigma_S(p_0)-\frac{1}{|\hat{G}|}G_A&=&Z_S^{-1}(p_0\gamma^0-a_S),\\
\Sigma_A(p_0)-\frac{1}{|\hat{G}|}G_S&=&Z_A^{-1}(p_0\gamma^0-a_A),
\end{eqnarray}
and $|\hat{G}|$ is defined by $|\hat{G}|=G_SG_A-v^2$.
The masses of $M_1$ and $M_2$ as functions of the mixing strength $v$
are shown for two cases in Fig.\ref{fig:massvsmixing}: for fixed
$G_1$ and $G_2$ (left) and for fixed $a_S$ and $a_A$ (right).

For both cases the mass difference is about several hundreds MeV, 
which is significantly smaller than the values of NQMs. 
The reason is that it is  generated by the mass
difference of the scalar and axial-vector diquark and their
mixing. The order of the amount is expected to be that of the spin-spin
interaction between the quarks. Hence the mass difference is almost the
same order as that of the nucleon and delta. 

For further extension of the present study, it is important to
investigate various transition amplitudes such as the decay rate and
 helicity amplitude of the Roper
resonance. These works are in progress.

This work is supported in part by the Grant for Scientific Research 
[(C) No.17959600]
from the Ministry of Education, Culture, Science and Technology,
Japan.


\begin{thebibliography}{99}

\bibitem{Nambu:1961fr}
  Y.~Nambu and G.~Jona-Lasinio,
  Phys.\ Rev.\  {\bf 124}, 246 (1961).

\bibitem{Ebert:1997hr}
  D.~Ebert and T.~Jurke,
  Phys.\ Rev.\ D {\bf 58}, 034001 (1998).

\bibitem{Abu-Raddad:2002pw}
  L.~J.~Abu-Raddad, A.~Hosaka, D.~Ebert and H.~Toki,
  Phys.\ Rev.\ C {\bf 66}, 025206 (2002).

\bibitem{Nagata:2004ky}
  K.~Nagata, A.~Hosaka and L.~J.~Abu-Raddad,
  Phys.\ Rev.\ C {\bf 72}, 035208 (2005).

\bibitem{Nagata:2003gg}
  K.~Nagata and A.~Hosaka,
  Prog.\ Theor.\ Phys.\  {\bf 111}, 857 (2004).

\bibitem{Jaffe:2003sg}
  R.~L.~Jaffe and F.~Wilczek,
  Phys.\ Rev.\ Lett.\  {\bf 91}, 232003 (2003).

\bibitem{Jaffe:2004ph}
  R.~L.~Jaffe,
  Phys.\ Rept.\  {\bf 409}, 1 (2005)
  [Nucl.\ Phys.\ Proc.\ Suppl.\  {\bf 142}, 343 (2005)].

\bibitem{Nagata:2005qb}
  K.~Nagata and A.~Hosaka,
  arXiv:hep-ph/0506193.

\bibitem{Ioffe:1981kw}
  B.~L.~Ioffe,
  Nucl.\ Phys.\ B {\bf 188}, 317 (1981)
  [Erratum-ibid.\ B {\bf 191}, 591 (1981)].

\end{thebibliography}
\end{document}